\documentclass{llncs}

\usepackage[german,english]{babel}
\usepackage{times,theorem}
\usepackage{latexsym,color,graphics}
\usepackage{diagrams}
\usepackage{url}
\usepackage{epsfig}
\usepackage{amssymb}
\usepackage{amsmath}
\usepackage{amsfonts}

\begingroup
\catcode`\~=11
\gdef\urltilde{\lower 0.6ex\hbox{~}}
\endgroup

\newcommand{\A}{\mathcal{A}} 
 \newcommand{\D}{\mathcal{D}}
\newcommand{\E}{\mathcal{E}}

 \renewcommand{\L}{\mathcal{L}}
\newcommand{\M}{\mathcal{M}} 
 \renewcommand{\P}{\mathcal{P}}
 \newcommand{\R}{\mathcal{R}}
 \newcommand{\T}{\mathcal{T}}
 \newcommand{\V}{\mathcal{V}}
\newcommand{\W}{\mathcal{W}}

\title{Intensionality and Two-steps Interpretations}
\author{Zoran Majki\'c}
\authorrunning{Zoran Majki\'c}
\institute{International Society for Research in Science and Technology \\
PO Box 2464 Tallahassee, FL 32316 - 2464 USA\\
\email{majk.1234@yahoo.com}\\ http://zoranmajkic.webs.com/}
\authorrunning{Zoran Majki\'c}

\newtheorem{propo}{Proposition}
\newtheorem{coro}{Corollary}

\newcount\pdfoutput
\begin{document}


\maketitle

\begin{abstract}
In this paper we considered the extension of the First-order Logic
(FOL) by Bealer's intensional abstraction operator. Contemporary use
of the term 'intension' derives from the traditional logical
Frege-Russell's doctrine that an idea (logic formula) has both an
extension and an intension. Although there is divergence in
formulation, it is accepted that the extension of an idea  consists
of the subjects to which the idea applies, and the intension
consists of the attributes implied by the idea. From the Montague's
point of view, the meaning of an idea  can be considered
as particular extensions in different possible worlds.\\
 In the case of
the pure FOL we obtain commutative homomorphic  diagram that holds
in each given possible world of the intensional FOL, from the free
algebra of the FOL syntax, toward its intensional algebra of
concepts, and, successively, to the new extensional relational
algebra (different from Cylindric algebras). Then we show that it
corresponds to the Tarski's interpretation of the standard
extensional FOL in this possible world.
\end{abstract}


\section{Introduction}
In "$\ddot{U}ber~ Sinn~ und~ Bedeutung$", Frege concentrated mostly
on the senses of names, holding that all names have a sense. It is
natural to hold that the same considerations apply to any expression
that has an extension. Two general terms can have the same extension
and different cognitive significance; two predicates can have the
same extension and different cognitive significance; two sentences
can have the same extension and different cognitive significance. So
general terms, predicates, and sentences all have senses as well as
extensions. The same goes for any expression that has an extension,
or is a candidate for extension.\\
The simplest aspect of an expression's meaning is its extension. We
can stipulate that the extension of a sentence is its truth-value,
and that the extension of a singular term is its referent. The
extension of other expressions can be seen as associated entities
that contribute to the truth-value of a sentence in a manner broadly
analogous to the way in which the referent of a singular term
contributes to the truth-value of a sentence. In many cases, the
extension of an expression will be what we intuitively think of as
its referent, although this need not hold in all cases, as the case
of sentences illustrates. While Frege himself is often interpreted
as holding that a sentence's referent is its truth-value, this claim
is counterintuitive and widely disputed. We can avoid that issue in
the present framework by using the technical term 'extension'. In
this context, the claim that the extension of a sentence is its
truth-value is a stipulation.\\
'Extensional' is most definitely a technical term. Say that the
extension of a name is its denotation, the extension of a predicate
is the set of things it applies to, and the extension of a sentence
is its truth value. A logic is extensional if coextensional
expressions can be substituted one for another in any sentence of
the logic "salva veritate", that is, without a change in truth
value. The intuitive idea behind this principle is that, in an
extensional logic, the only logically significant notion of meaning
that attaches to an expression is its extension. An intensional
logics is exactly one in which substitutivity salva veritate fails
for some of the sentences of the logic.\\
 The first conception of intensional entities (or concepts) is built
into the \emph{possible-worlds} treatment of Properties, Relations
and Propositions (PRP)s. This conception is commonly attributed to
Leibniz, and underlies Alonzo Church's alternative formulation of
Frege's theory of senses ("A formulation of the Logic of Sense and
Denotation" in Henle, Kallen, and Langer, 3-24, and "Outline of a
Revised Formulation of the Logic of Sense and Denotation" in two
parts, Nous,VII (1973), 24-33, and VIII,(1974),135-156). This
conception of PRPs is ideally suited for treating the
\emph{modalities} (necessity, possibility, etc..) and to Montague's
definition of intension of a given virtual predicate
$\phi(x_1,...,x_k)$ (a FOL open-sentence with the tuple of free
variables $(x_1,...x_k)$) as a mapping from possible worlds into
extensions of this virtual predicate. Among the possible worlds we
distinguish the \emph{actual} possible world. For example if we
consider a set of predicates of a given Database
and their extensions in different time-instances, the actual possible world is identified by the current instance of the time.\\
The second conception of intensional entities is to be found in in
Russell's doctrine of logical atomism. On this doctrine it is
required that all complete definitions of intensional entities be
finite as well as unique and non-circular: it offers an
\emph{algebraic} way for definition of complex intensional entities
from simple (atomic) entities (i.e., algebra of concepts),
conception also evident in Leibniz's remarks. In a predicate logics,
predicates and open-sentences (with free variables) expresses
classes (properties and relations), and sentences express
propositions. Note that classes (intensional entities) are
\emph{reified}, i.e., they belong to the same domain as individual
objects (particulars). This endows the intensional logics with a
great deal of uniformity, making it possible to manipulate classes
and individual objects in the same language. In particular, when
viewed as an individual
object, a class can be a member of another class.\\
 The distinction between
intensions and extensions is important, considering that extensions
can be notoriously difficult to handle in an efficient manner. The
extensional equality theory of predicates and functions under
higher-order semantics (for example, for two predicates with the
same set of attributes $p = q$ is true iff these symbols are
interpreted by the same relation), that is, the strong equational
theory of intensions, is not decidable, in general. For example, in
the second-order predicate calculus and Church's simple theory of
types, both under the standard semantics, is not even
semi-decidable. Thus, separating intensions from extensions makes it
possible to have an equational theory over predicate and function
names (intensions) that is separate from the extensional equality of
relations and functions. \\
In what follows we  denote by $B^A$ the set of all functions from
$A$ to $B$, and by $A^n$ a n-folded cartesian product $A \times
...\times A$ for $n \geq 1$. By $f, t$ we denote  empty set
$\emptyset$ and singleton set $\{<>\}$ respectively (with the empty
tuple $<>$ i.e. the unique tuple of 0-ary relation), which may be
thought of
 as falsity $f$ and truth $t$, as those used  in the relational algebra.
 For a given domain $\D$ we define that $\D^0$ is a singleton set $\{<>\}$, so that $\{f, t\} = \P(\D^0)$, where $\P$ is the powerset operator.
\section{Intensional FOL language with intensional abstraction} \label{section:intensional}
%
 Intensional entities are such concepts as
propositions and properties. What make them 'intensional' is that
they violate the principle of extensionality; the principle that
extensional equivalence implies identity. All (or most) of these
intensional entities have been classified at one time or another as
kinds of Universals \cite{Beal93}.\\
We consider a non empty domain $~\D = D_{-1} \bigcup D_I$,  where a
subdomain $D_{-1}$ is made of
 particulars (extensional entities), and the rest $D_I = D_0 \bigcup
 D_1 ...\bigcup D_n ...$ is made of
 universals ($D_0$ for propositions (the 0-ary concepts), and  $D_n, n \geq 1,$ for
 n-ary concepts.\\
 The fundamental entities
are \emph{intensional abstracts} or so called, 'that'-clauses. We
assume that they are singular terms; Intensional expressions like
'believe', mean', 'assert', 'know',
 are standard two-place predicates  that take 'that'-clauses as
 arguments. Expressions like 'is necessary', 'is true', and 'is
 possible' are one-place predicates that take 'that'-clauses as
 arguments. For example, in the intensional sentence "it is
 necessary that $\phi$", where $\phi$ is a proposition, the 'that $\phi$' is
 denoted by the $\lessdot \phi \gtrdot$, where $\lessdot \gtrdot$ is the intensional abstraction
 operator which transforms a logic formula into a \emph{term}. Or, for example, "x believes that $\phi$" is given by formula
$p_i^2(x,\lessdot \phi \gtrdot)$ ( $p_i^2$ is binary 'believe'
predicate).\\
Here we will present an  intensional FOL  with slightly different
intensional abstraction than that originally presented  in
\cite{Beal79}:
 \begin{definition} \label{def:bealer}
  The syntax of the First-order Logic language with intensional abstraction
$\lessdot \gtrdot$, denoted by $\L$, is as follows:\\
 Logic operators $(\wedge, \neg, \exists)$; Predicate letters in $P$
 (functional letters are considered as particular case of predicate
 letters); Variables $x,y,z,..$ in $\V$; Abstraction $\lessdot \_ \gtrdot$, and punctuation
 symbols (comma, parenthesis).
 With the following simultaneous inductive definition of \emph{term} and
 \emph{formula}   :\\
   1. All variables and constants (0-ary functional letters in P) are terms.\\
   2. If $~t_1,...,t_k$ are terms, then $p_i^k(t_1,...,t_k)$ is a formula
 ($p_i^k \in P$ is a k-ary predicate letter).\\
   3. If $\phi$ and $\psi$ are formulae, then $(\phi \wedge \psi)$, $\neg \phi$, and
 $(\exists x)\phi$ are formulae. \\
   4. If $\phi(\textbf{x})$ is a formula (virtual predicate) with a list of free variables in $\textbf{x} =(x_1,...,x_n)$ (with ordering
from-left-to-right of their appearance in $\phi$), and  $\alpha$ is
its sublist of \emph{distinct} variables,
 then $\lessdot \phi \gtrdot_{\alpha}^{\beta}$ is a term, where $\beta$ is the remaining list of free variables preserving ordering in $\textbf{x}$ as well. The externally quantifiable variables are the \emph{free} variables not in $\alpha$.
  When $n =0,~ \lessdot \phi \gtrdot$ is a term which denotes a
proposition, for $n \geq 1$ it denotes
 a n-ary concept.\\
An occurrence of a variable $x_i$ in a formula (or a term) is
\emph{bound} (\emph{free}) iff it lies (does not lie) within a
formula of the form $(\exists x_i)\phi$ (or a term of the form
$\lessdot \phi \gtrdot_{\alpha}^{\beta}$ with $x_i \in \alpha$). A
variable is free (bound) in a formula (or term) iff it has (does not
have) a free occurrence in that formula (or term).\\ A
\emph{sentence} is a formula having no free variables. The binary
predicate letter $p_1^2$ for identity is singled out as a
distinguished logical predicate and formulae of the form
$p^2_1(t_1,t_2)$ are to be rewritten in the form $t_1 \doteq t_2$.
We denote by $R_{=}$ the binary relation obtained by standard
Tarski's interpretation of this predicate $p^2_1$. The logic
operators $\forall, \vee, \Rightarrow$ are defined in terms of
$(\wedge, \neg, \exists)$ in the usual way.
\end{definition}
The universal quantifier is defined by $\forall = \neg \exists
\neg$. Disjunction and implication are expressed by
 $\phi \vee \psi = \neg(\neg \phi \wedge \neg \psi)$, and $\phi \Rightarrow \psi = \neg \phi \vee
 \psi$.
 In FOL with the
identity $\doteq$, the formula $(\exists_1 x)\phi(x)$ denotes the
formula $(\exists x)\phi(x) \wedge (\forall x)(\forall y)(\phi(x)
\wedge \phi(y)  \Rightarrow (x \doteq y))$. We denote by $R_{=}$ the
Tarski's interpretation of
$\doteq$.\\
In what follows any open-sentence, a formula $\phi$ with non empty
tuple of free variables $(x_1,...,x_m)$, will be called a m-ary
  \emph{virtual predicate}, denoted also by
$\phi(x_1,...,x_m)$. This definition contains the precise method of
establishing the \emph{ordering} of variables in this tuple:
  such an method that will be adopted here is the ordering of appearance, from left to right, of free variables in $\phi$.
   This method of composing the tuple of free variables
  is the unique and canonical way of definition of the virtual predicate from a given
  formula.\\
The \emph{intensional interpretation} of this intensional FOL is a
mapping between the set $\L$ of formulae of the logic language  and
 intensional entities in $\D$, $I:\L \rightarrow \D$, is a kind of
 "conceptualization", such that  an open-sentence (virtual
 predicate)
 $\phi(x_1,...,x_k)$ with a tuple of all free variables
 $(x_1,...,x_k)$ is mapped into a k-ary \emph{concept}, that is, an intensional entity  $u =
 I(\phi(x_1,...,x_k)) \in D_k$, and (closed) sentence $\psi$ into a proposition (i.e., \emph{logic} concept) $v =
 I(\psi) \in D_0$ with $I(\top) = Truth \in D_0$ for the FOL tautology $\top$. A language constant $c$ is mapped into a
 particular $a = I(c) \in D_{-1}$ if it is a proper name, otherwise in a correspondent concept in
$\D$. \\
 An assignment $g:\V \rightarrow \D$ for variables in $\V$ is
applied only to free variables in terms and formulae.  Such an
assignment $g \in \D^{\V}$ can be recursively uniquely extended into
the assignment $g^*:\T \rightarrow \D$, where $\T$ denotes the set
of all terms (here $I$ is an intensional interpretation of this FOL,
as explained
in what follows), by :\\
1. $g^*(t) = g(x) \in \D$ if the term $t$ is a variable $x \in
\V$.\\
2. $g^*(t) = I(c) \in \D$ if the term $t$ is a constant $c \in
P$.\\
3. if $t$ is an abstracted term $\lessdot \phi
\gtrdot_{\alpha}^{\beta}$,  then $g^*(\lessdot \phi
\gtrdot_{\alpha}^{\beta}) = I(\phi[\beta /g(\beta)] ) \in D_k, k =
|\alpha|$ (i.e., the number of variables in $\alpha$), where
$g(\beta) = g(y_1,..,y_m) = (g(y_1),...,g(y_m))$ and $[\beta
/g(\beta)]$ is a uniform replacement of each i-th variable in the
list $\beta$
with the i-th constant in the list $g(\beta)$. Notice that $\alpha$ is the list of all free variables in the formula $\phi[\beta /g(\beta)]$.\\
  We denote by $~t/g~$ (or $\phi/g$) the ground term (or
formula) without free variables, obtained by assignment $g$ from a
term $t$ (or a formula $\phi$), and by  $\phi[x/t]$ the formula
obtained by  uniformly replacing $x$ by a term $t$ in $\phi$.\\
The distinction between intensions and extensions is important
 especially because we are now able to have and \emph{equational
 theory} over intensional entities (as  $\lessdot \phi \gtrdot$), that
 is predicate and function "names", that is separate from the
 extensional equality of relations and functions.
 An extensionalization function $h$ assigns to the intensional elements of $\D$ an appropriate
extension as follows: for each proposition $u \in D_0$, $h(u) \in
 \{f,t\} \subseteq \P(D_{-1})$ is its
 extension (true or false value); for each n-ary
 concept $u \in D_n$, $h(u)$ is a subset of $\D^n$
 (n-th Cartesian product of $\D$); in the case of particulars $u \in
 D_{-1}$, $h(u) = u$.\\
The sets $f, t$
   are empty set $\{\}$ and set $\{<>\}$ (with the empty tuple $<> \in D_{-1}$ i.e. the unique tuple of 0-ary relation)
 which may be thought of
 as falsity and truth, as those used  in the Codd's relational-database algebra \cite{Codd70} respectively,
 while $Truth \in D_0$ is the concept (intension)
of the tautology. \\
 We define that $\D^0 = \{<>\}$, so that $\{f,t\} = \P(\D^0)$.
 Thus we have:
 \\$h = h_{-1}  + \sum_{i\geq 0}h_i:\sum_{i
\geq -1}D_i \longrightarrow D_{-1} +  \sum_{i\geq 0}\P(D^i)$,\\
 where $h_{-1} = id:D_{-1} \rightarrow D_{-1}$
is identity, $h_0:D_0 \rightarrow \{f,t\}$ assigns truth values in $
\{f,t\}$, to all propositions, and $h_i:D_i \rightarrow \P(D^i)$,
$i\geq 1$, assigns extension to all concepts, where $\P$ is the
powerset operator. Thus, intensions can be seen as \emph{names} of
abstract or concrete entities, while extensions correspond to
various rules that these entities play in different worlds.\\
\textbf{Remark:} (Tarski's constraint) This semantics has to
preserve Tarski's semantics of the FOL, that is, for any formula
$\phi \in \L$ with the tuple of free variables $(x_1,...,x_k)$, any
assignment $g \in \D^{\V}$, and every $h \in \E$ it has to be
satisfied that: \\ (T)$~~~h(I(\phi/g)) = t~~$ iff
$~~(g(x_1),...,g(x_k)) \in h(I(\phi))$.\\
$\square$\\
 Thus, intensional  FOL  has the simple Tarski
first-order semantics, with a decidable
 unification problem, but we need also the actual world mapping
 which maps any intensional entity to its \emph{actual world
 extension}. In what follows we will identify a \emph{possible world} by a
 particular mapping which assigns to intensional entities their
 extensions in such possible world. That is direct bridge between
 intensional FOL  and possible worlds representation
 \cite{Lewi86,Stal84,Mont70,Mont73,Mont74,Majk09FOL}, where intension of a proposition is a
 \emph{function} from possible worlds $\W$ to truth-values, and
 properties and functions from $\W$ to sets of possible (usually
 not-actual) objects.\\
 Here $\E$ denotes the set of possible
\emph{extensionalization functions} that satisfy the constraint (T);
they can be considered as \emph{possible worlds} (as in Montague's
intensional semantics for natural language \cite{Mont70,Mont74}), as
demonstrated in \cite{Majk08in,Majk08ird}, given by the bijection
$~~~is:\W \simeq \E$.\\
Now we are able to define formally this intensional semantics
\cite{Majk09FOL}:
 \begin{definition} \label{def:intensemant} \textsc{Two-step \textsc{I}ntensional
 \textsc{S}emantics:}\\
Let $\mathfrak{R} = \bigcup_{k \in \mathbb{N}} \P(\D^k) = \sum_{k\in
\mathbb{N}}\P(D^k)$ be the set of all k-ary relations, where $k \in
\mathbb{N} = \{0,1,2,...\}$. Notice that $\{f,t\} = \P(\D^0) \in
\mathfrak{R}$, that is, the truth values are extensions in
$\mathfrak{R}$.\\ The intensional semantics of the logic language
with the set of formulae $\L$ can be represented by the  mapping
\begin{center}
$~~~ \L ~\longrightarrow_I~ \D ~\Longrightarrow_{w \in \W}~
\mathfrak{R}$,
\end{center}
where $~\longrightarrow_I~$ is a \emph{fixed intensional}
interpretation $I:\L \rightarrow \D$ and $~\Longrightarrow_{w \in
\W}~$ is \emph{the set} of all extensionalization functions $h =
is(w):\D \rightarrow \mathfrak{R}$ in $\E$, where $is:\W \rightarrow
\E$ is the mapping from the set of possible worlds to the set of
 extensionalization functions.\\
 We define the mapping $I_n:\L_{op} \rightarrow
\mathfrak{R}^{\W}$, where $\L_{op}$ is the subset of formulae with
free variables (virtual predicates), such that for any virtual
predicate $\phi(x_1,...,x_k) \in \L_{op}$ the mapping
$I_n(\phi(x_1,...,x_k)):\W \rightarrow \mathfrak{R}$ is the
Montague's meaning (i.e., \emph{intension}) of this virtual
predicate \cite{Lewi86,Stal84,Mont70,Mont73,Mont74}, that is, the
mapping which returns with the extension of this (virtual) predicate
in every possible world in $\W$.
\end{definition}
We adopted this two-step intensional semantics, instead of well
known Montague's semantics (which lies in the construction of a
compositional and recursive semantics that covers both intension and
extension) because of a number of its weakness.\\
\textbf{Example}:  Let us consider the following two past
participles: 'bought' and 'sold'(with unary predicates $p_1^1(x)$,
'$x$ has been bought', and $p_2^1(x)$,'$x$ has been sold'). These
two different concepts in the Montague's semantics would have not
only the same extension but also their intension, from the fact that
their extensions are identical in every possible world.\\ Within the
two-steps formalism we can avoid this problem by assigning two
different concepts (meanings) $u = I(p_1^1(x))$ and $ v =
I(p_2^1(x))$ in $\in D_1$. Notice that the same problem we have in
the Montague's semantics for two sentences with different meanings,
which bear the same truth value across all possible worlds: in the
Montague's semantics they will be forced to the \emph{same}
meaning.\\$\square$\\
 Another relevant question w.r.t. this two-step
interpretations of an intensional semantics is how in it is managed
the extensional identity relation $\doteq$ (binary predicate of the
identity) of the FOL. Here this extensional identity relation is
mapped into the binary concept $Id = I(\doteq(x,y)) \in D_2$, such
that $(\forall w \in \W)(is(w)(Id) = R_{=})$, where $\doteq(x,y)$
(i.e., $p_1^2(x,y)$) denotes an atom of the FOL of the binary
predicate for identity in FOL, usually written by FOL formula $x
\doteq y$ (here we prefer to distinguish this \emph{formal symbol}
$~ \doteq ~ \in P$ of the built-in identity binary predicate letter
in the FOL from the standard mathematical
symbol '$=$' used in all mathematical definitions in this paper).\\
 In what follows we will use the function $f_{<>}:\mathfrak{R}
\rightarrow \mathfrak{R}$, such that for any $R \in \mathfrak{R}$,
$f_{<>}(R) = \{<>\}$ if $R \neq \emptyset$; $\emptyset$ otherwise.
Let us define the following set of algebraic operators for
 relations in $\mathfrak{R}$:
\begin{enumerate}
\item binary operator $~\bowtie_{S}:\mathfrak{R} \times \mathfrak{R} \rightarrow
\mathfrak{R}$,
 such that for any two relations $R_1, R_2 \in
 \mathfrak{R}~$, the
 $~R_1 \bowtie_{S} R_2$ is equal
to the relation obtained by natural join
 of these two relations $~$ \verb"if"
 $S$ is a non empty
set of pairs of joined columns of respective relations (where the
first argument is the column index of the relation $R_1$ while the
second argument is the column index of the joined column of the
relation $R_2$); \verb"otherwise" it is equal to the cartesian
product $R_1\times R_2$. For example, the logic formula
$\phi(x_i,x_j,x_k,x_l,x_m) \wedge \psi (x_l,y_i,x_j,y_j)$ will be
traduced by the algebraic expression $~R_1 \bowtie_{S}R_2$ where
$R_1 \in \P(\D^5), R_2\in \P(\D^4)$ are the extensions for a given
Tarski's interpretation  of the virtual predicate $\phi, \psi$
relatively, so that $S = \{(4,1),(2,3)\}$ and the resulting relation
will have the following ordering of attributes:
$(x_i,x_j,x_k,x_l,x_m,y_i,y_j)$.
\item unary operator $~ \sim:\mathfrak{R} \rightarrow \mathfrak{R}$, such that for any k-ary (with $k \geq 0$)
relation $R \in  \P(\D^{k}) \subset \mathfrak{R}$
 we have that $~ \sim(R) = \D^k \backslash R \in \D^{k}$, where '$\backslash$' is the substraction of relations. For example, the
logic formula $\neg \phi(x_i,x_j,x_k,x_l,x_m)$ will be traduced by
the algebraic expression $~\D^5 \backslash R$ where $R$ is the
extensions for a given Tarski's interpretation  of the virtual
predicate $\phi$.
\item unary operator $~ \pi_{-m}:\mathfrak{R} \rightarrow \mathfrak{R}$, such that for any k-ary (with $k \geq 0$) relation $R \in \P(\D^{k}) \subset \mathfrak{R}$
we have that $~ \pi_{-m} (R)$ is equal to the relation obtained by
elimination of the m-th column of the relation $R~$ \verb"if" $1\leq
m \leq k$ and $k \geq 2$; equal to $~f_{<>}(R)~$ \verb"if" $m = k
=1$; \verb"otherwise" it is equal to $R$. For example, the logic
formula $(\exists x_k) \phi(x_i,x_j,x_k,x_l,x_m)$ will be traduced
by the algebraic expression $~\pi_{-3}(R)$ where $R$ is the
extensions for a given Tarski's interpretation  of the virtual
predicate $\phi$ and the resulting relation will have the following
ordering of attributes: $(x_i,x_j,x_l,x_m)$.
\end{enumerate}
Notice that the ordering of attributes of resulting relations
corresponds to the method used for generating the ordering of
variables in the tuples of free variables adopted for virtual
predicates.\\
  Analogously to Boolean algebras
 which are extensional models of propositional logic, we introduce an
 intensional algebra for this intensional FOL as follows.
\begin{definition}  \label{def:int-algebra} Intensional algebra for the intensional FOL  in Definition \ref{def:bealer} is a structure $~\A_{int}
= ~(\D,  f, t, Id, Truth,  \{conj_{S}\}_{ S \in \P(\mathbb{N}^2)},
neg,
\\\{exists_{n}\}_{n \in \mathbb{N}})$,  $~~$  with
 binary operations  $~~conj_{S}:D_I\times D_I \rightarrow D_I$,
   unary operation  $~~neg:D_I\rightarrow D_I$,  unary
   operations $~~exists_{n}:D_{I}\rightarrow D_I$,  such that for any
extensionalization function $h \in \E$,
and $u \in D_k, v \in D_j$, $k,j \geq 0$,\\
1. $~h(Id) = R_=~$ and $~h(Truth) = \{<>\}$.\\
2. $~h(conj_{S}(u, v)) = h(u) \bowtie_{S}h(v)$, where $\bowtie_{S}$
is the natural join operation defined above and $conj_{S}(u, v) \in
D_m$ where $m = k + j - |S|$
 if for every pair $(i_1,i_2) \in S$ it holds that $1\leq i_1 \leq k$, $1 \leq i_2 \leq j$ (otherwise $conj_{S}(u, v) \in D_{k+j}$).\\
3. $~h(neg(u)) = ~\sim(h(u)) = \D^k \backslash (h(u))$,
 where  $~\sim~$ is the operation
defined above and $neg(u) \in D_k$.\\
 4. $~h(exists_{n}(u)) =
\pi_{-n}(h(u))$, where $\pi_{-n}$ is the operation defined above and
\\ $exists_n(u) \in D_{k-1}$ if $1 \leq n \leq k$ (otherwise
$exists_n$ is the identity function).
\end{definition}
Notice that for $u \in D_0$, $~h(neg(u)) = ~\sim(h(u)) = \D^0
\backslash (h(u)) = \{<>\} \backslash (h(u)) \in \{f,t\}$.\\ We
define a derived operation $~~union:(\P(D_i)\backslash \emptyset)
\rightarrow D_i$, $i \geq 0$, such that, for any $B =
\{u_1,...,u_n\} \in \P(D_i)$ we have that
$union(\{u_1,...,u_n\})=_{def} ~u_1$ if $n = 1$; $
neg(conj_S(neg(u_1),conj_S(...,neg(u_n))...)$, where $S =
\{(l,l)~|~1 \leq l \leq i \}$, otherwise. Than we obtain that for $n \geq 2$:\\
$h(union (B) = h(neg(conj_S(neg(u_1),conj_S(...,neg(u_n))...) \\
 = \D^i\backslash((\D^i\backslash h(u_1)
\bowtie_{S}...\bowtie_{S}(\D^i\backslash h(u_n)) \\
= \D^i\backslash((\D^i\backslash h(u_1)
\bigcap...\bigcap(\D^i\backslash h(u_n)) \\
= \bigcup\{ h(u_j)~|~1 \leq j \leq n\} =  ~\bigcup \{h(u) ~|~u \in
B\}$. \\
Intensional interpretation $I:\L \rightarrow \D$ satisfies the
following homomorphic extension:
\begin{enumerate}
  \item The logic formula $\phi(x_i,x_j,x_k,x_l,x_m) \wedge \psi
(x_l,y_i,x_j,y_j)$ will be intensionally interpreted by the concept
$u_1 \in D_7$, obtained by the algebraic expression $~
conj_{S}(u,v)$ where $u = I(\phi(x_i,x_j,x_k,x_l,x_m)) \in D_5, v =
I(\psi (x_l,y_i,x_j,y_j))\in D_4$ are the concepts of the virtual
predicates $\phi, \psi$, relatively, and $S = \{(4,1),(2,3)\}$.
Consequently, we have that for any two formulae $\phi,\psi \in \L$
and a particular  operator $conj_S$ uniquely determined by tuples of
free variables in these two formulae, $I(\phi \wedge \psi) =
conj_{S}(I(\phi),I(\psi))$.
  \item The logic formula $\neg \phi(x_i,x_j,x_k,x_l,x_m)$ will be
intensionally interpreted by the concept $u_1  \in D_5$, obtained by
the algebraic expression $~neg(u)$ where $u =
I(\phi(x_i,\\x_j,x_k,x_l,x_m)) \in D_5$ is the concept of the
virtual predicate $\phi$. Consequently, we have that for any formula
$\phi \in \L$, $~I(\neg \phi) = neg(I(\phi))$.
  \item The logic formula $(\exists x_k) \phi(x_i,x_j,x_k,x_l,x_m)$ will
be intensionally interpreted by the concept $u_1  \in D_4$, obtained
by the algebraic expression $~exists_{3}(u)$ where $u =
I(\phi(x_i,x_j,x_k,x_l,x_m)) \in D_5$ is the concept of the virtual
predicate $\phi$. Consequently, we have that for any formula $\phi
\in \L$ and a particular operator $exists_{n}$ uniquely determined
by the position of the  existentially quantified variable in the
tuple of free variables in $\phi$ (otherwise $n =0$ if this
quantified variable is not a free variable in $\phi$), $~I((\exists
x)\phi) = exists_{n}(I(\phi))$.
\end{enumerate}
 Once one has found a method for specifying the interpretations of
singular terms of $\L$ (take in consideration the particularity of
abstracted terms), the Tarski-style definitions of truth and
validity for  $\L$ may be given in the customary way.
What is being south specifically is a method for characterizing the
intensional interpretations of singular terms of $\L$ in such a way
that a given singular abstracted term $\lessdot \phi
\gtrdot_{\alpha}^{\beta}$ will denote an appropriate property,
relation, or proposition, depending on the value of $m = |\alpha|$.
Thus, the mapping of intensional abstracts (terms)  into $\D$ we
will define differently from that given in the version of Bealer
\cite{Beal82}, as follows:
\begin{definition}  \label{def:abstraction} An intensional
interpretation $I$ can be extended to abstracted terms as follows:
for any abstracted term $\lessdot \phi \gtrdot_{\alpha}^{\beta}$  we
define that,\\
$I(\lessdot \phi \gtrdot_{\alpha}^{\beta} ) = union (\{I(\phi[\beta/
g(\beta)])~|~ g \in \D^{\overline{\beta}}\})$,\\
where $\overline{\beta}$ denotes the \verb"set" of elements in the
list $\beta$, and the assignments in  $\D^{\overline{\beta}}$ are
limited only to the variables in $\overline{\beta}$.
 \end{definition}
\textbf{Remark:} Here we can make the question if there is a sense
to extend the interpretation also to (abstracted) terms, because in
Tarski's interpretation of FOL we do not have any interpretation for
terms, but only the assignments for terms, as we defined previously
by the mapping $g^*:\T \rightarrow\D$. The answer is positive,
because the abstraction symbol $\lessdot \_~
\gtrdot_{\alpha}^{\beta}$ can be considered as a kind of the unary
built-in functional symbol of intensional FOL, so that we can apply
the Tarskian interpretation to this functional symbol into the fixed
mapping $I(\lessdot \_~ \gtrdot_{\alpha}^{\beta}):\L \rightarrow
\D$, so that for any $\phi \in \L$ we have that $I(\lessdot \phi
\gtrdot_{\alpha}^{\beta})$ is equal to the application of this
function to the value $\phi$, that is, to $I(\lessdot \_~
\gtrdot_{\alpha}^{\beta})(\phi)$. In such an approach we would
introduce also the typed variable $X$ for the formulae in $\L$, so
that the Tarskian assignment for this
functional symbol with variable $X$, with $g(X) = \phi \in \L$, can be given by:\\
$g^*(\lessdot \_~ \gtrdot_{\alpha}^{\beta}(X)) = I(\lessdot \_~
\gtrdot_{\alpha}^{\beta})(g(X)) = I(\lessdot \_~
\gtrdot_{\alpha}^{\beta})(\phi)\\ = ~<>\in D_{-1},~$ if
$~\overline{\alpha} \bigcup \overline{\beta}$ is not
equal to the set of free variables in $\phi$;\\
$= union (\{I(\phi[\beta/ g'(\beta)])~|~ g' \in
\D^{\overline{\beta}}\}) \in D_{|\overline{\alpha}|}$, otherwise.
\\$\square$\\
 Notice than if $\beta = \emptyset$ is the empty
list, then $I(\lessdot \phi \gtrdot_{\alpha}^{\beta} ) = I(\phi)$.
Consequently, the denotation of $\lessdot \phi\gtrdot $
 is equal to the meaning of a proposition $\phi$, that is, $~I(\lessdot \phi\gtrdot) =
I(\phi)\in D_0$.  In the case when $\phi$ is an atom
$p^m_i(x_1,..,x_m)$ then $I (\lessdot
p^m_i(x_1,..,x_m)\gtrdot_{x_1,..,x_m}) = I(p^m_i(x_1,..,x_m)) \in
D_m$, while \\$I (\lessdot p^m_i(x_1,..,x_m)\gtrdot^{x_1,..,x_m}) =
union (\{I(p^m_i(g(x_1),..,g(x_m)))~|~ g \in \D^{\{x_1,..,x_m\}} \})
\in D_0$,  with $h(I (\lessdot
p^m_i(x_1,..,x_m)\gtrdot^{x_1,..,x_m})) = h(I((\exists
x_1)...(\exists x_m)p^m_i(x_1,..,x_m))) \in \{f,t\}$.\\ For
example,\\ $h(I(\lessdot p^1_i(x_1) \wedge \neg p^1_i(x_1)
\gtrdot^{x_1})) = h(I((\exists x_1)(\lessdot p^1_i(x_1) \wedge \neg
p^1_i(x_1)
\gtrdot^{x_1}))) = f$.\\
The interpretation of a more complex abstract $\lessdot \phi
\gtrdot_\alpha^{\beta}$ is defined in terms of the interpretations
of the relevant syntactically simpler expressions, because the
interpretation of more complex formulae is defined in terms of the
interpretation of the relevant syntactically simpler formulae, based
on the intensional algebra above. For example, $I(p_i^1(x) \wedge
p_k^1(x)) = conj_{\{(1,1)\}}(I(p_i^1(x)), I(p_k^1(x)))$, $I(\neg
\phi) = neg(I(\phi))$, $I(\exists x_i)\phi(x_i,x_j,x_i,x_k) = exists_3(I(\phi))$.\\
Consequently, based on the intensional algebra in Definition
\ref{def:int-algebra} and on intensional interpretations of
abstracted term in Definition \ref{def:abstraction}, it holds that
the interpretation of any formula in $\L$ (and any abstracted term)
will be reduced to an algebraic expression over interpretation of
primitive atoms in $\L$. This obtained expression is finite for any
finite formula (or abstracted term), and represents the \emph{
meaning} of such finite formula (or abstracted term).\\
The \emph{extension} of abstracted terms satisfy the following
property:
\begin{propo}  \label{prop:abstraction}
For any abstracted term $\lessdot \phi \gtrdot_{\alpha}^{\beta}$
with $|\alpha| \geq 1$ we have that \\$~~ h(I(\lessdot \phi
\gtrdot_{\alpha}^{\beta} )) = \pi_{- \beta}(h(I(\phi)))$,\\ where
$\pi_{- (y_1,...,y_k)} = \pi_{-y_1} \circ ...\circ \pi_{-y_1}$,
$\circ$ is the sequential composition of functions), and  $\pi_{-
\emptyset}$ is an identity.
\end{propo}
\textbf{Proof:} Let $\textbf{x}$ be a tuple of all free variables in
$\phi$, so that $\overline{\textbf{x}} = \overline{\alpha} \bigcup
\overline{\beta}$, $\alpha = (x_1,...,x_k)$, then we have that
$~~ h(I(\lessdot \phi \gtrdot_{\alpha}^{\beta} )) =\\
 = h(union (\{I(\phi[\beta/
g(\beta)])~|~ g \in \D^{\overline{\beta}}\}))$, from Def.
\ref{def:abstraction}\\
$= \bigcup \{h(I(\phi[\beta/ g(\beta)]))~|~ g \in
\D^{\overline{\beta}}\}$ \\
$= \bigcup \{ \{(g_1(x_1),...,g_1(x_k))~|~ g_1 \in
\D^{\overline{\alpha}}$ and $ h(I(\phi[\beta/ g(\beta)][\alpha/
g_1(\alpha)])) = t\}~|~ g \in \D^{\overline{\beta}}\}$\\
$=  \{g_1(\alpha)~|~ g_1 \in \D^{\overline{\alpha}\bigcup
\overline{\beta}}$ and $ h(I(\phi/g_1))
= t\}\\
=  \pi_{- \beta}(\{g_1(\textbf{x})~|~ g_1 \in
\D^{\overline{\textbf{x}}}$ and $ h(I(\phi/g_1))
= t\})\\
=  \pi_{- \beta}(\{g_1(\textbf{x})~|~ g_1 \in
\D^{\overline{\textbf{x}}}$ and $g_1(\textbf{x})\in h(I(\phi))
\})$, by (T) \\
$ =  \pi_{- \beta}(h(I(\phi)))$.
\\$\square$\\
 We can connect $\E$
 with a possible-world semantics. Such a correspondence is a natural identification of
 intensional logics with modal Kripke based logics.
\begin{definition} (Model): \label{def:Semant} A model for  intensional FOL with fixed
intensional interpretation $I$, which express the two-step
intensional semantics in Definition \ref{def:intensemant}, is the
Kripke structure $\M_{int} = (\W, \D, V)$, where $\W =
\{is^{-1}(h)~|~h \in \E\}$,  a mapping $~V:\W \times P \rightarrow
\bigcup_{n < \omega} \{t,f\}^{\D^n}$, with $P$ a set of predicate
symbols of the language, such that for any world $w = is^{-1}(h) \in
\W, p^n_i \in P$, and $(u_1,...,u_n) \in \D^n$ it holds that
$V(w,p^n_i)(u_1,...,u_n) = h(I(p^n_i(u_1,...,u_n)))$. The
satisfaction relation $\models_{w,g}$ for a given $w \in \W$ and
assignment $g \in \D{\V}$ is defined as follows:\\
1. $~{\M} \models_{w,g}p_i^k(x_1,...,x_k)~$ iff $~V(w,p_i^k)(g(x_1),...,g(x_k)) = t$,\\
 2. $~{\M} \models_{w,g} \varphi \wedge \phi~$ iff $~{\M} \models_{w,g} \varphi~$ and $~{\M} \models_{w,g}
 \phi$, \\
 3. $~{\M} \models_{w,g} \neg \varphi ~$ iff $~$ not ${\M} \models_{w,g}
\varphi$,\\
 4. $~\M \models_{w,g}
(\exists x) \phi ~~$ iff \\
4.1. $~\M \models_{w,g} \phi$, if $x$ is not a free
variable in $\phi$;\\
4.2. $~$exists $u \in \D$ such that $~\M \models_{w,g} \phi[x/u]$,
if $x$ is  a free variable in $\phi$.
 \end{definition}
It is easy to show that  the satisfaction relation $\models$ for
this Kripke semantics in a world $w = is^{-1}(h)$ is defined
by, $~~\M \models_{w,g} \phi~~$ iff $~~h(I(\phi/g)) = t$.\\
We can enrich this intensional FOL by another modal operators, as
for example the "necessity" universal operator $\Box$ with an
accessibility relation $~~\R =\W \times \W$, obtaining the S5 Kripke
structure $\M_{int} = (\W, \R, \D, V)$, in order to be able to
define the following equivalences between the abstracted terms
without free variables $\lessdot \phi \gtrdot_{\alpha}^{\beta_1}/g$
and $\lessdot \psi \gtrdot_{\alpha}^{\beta_2}/g$, where all free
variables (not in $\alpha$) are instantiated by $g \in \D^{\V}$
(here $A \equiv B$ denotes the formula $(A \Rightarrow B) \wedge (B
\Rightarrow A)$):
\begin{itemize}
  \item (Strong) intensional
equivalence (or \emph{equality}) "$\asymp$"  is defined by:\\
$\lessdot \phi \gtrdot_{\alpha}^{\beta_1}/g ~\asymp~  \lessdot \psi
\gtrdot_{\alpha}^{\beta_2}/g~~~~~$
iff $~~~~~\Box ( \phi[\beta_1/g(\beta_1)] \equiv \psi[\beta_2/g(\beta_2)])$, \\
 with $~{\M} \models_{w,g'}~\Box \varphi~$ iff $~$ for all $w'\in \W$, $(w,w')
\in {\R}$ implies ${\M} \models_{w',g'}\varphi$.\\
From Example 1 we have that $\lessdot p_1^1(x)\gtrdot_{x} ~\asymp~
\lessdot p_2^1(x) \gtrdot_{x}$, that is '$x$ has been bought' and
'$x$ has been sold' are intensionally equivalent, but they have not
the same meaning (the concept $I(p_1^1(x)) \in D_1$ is different
from $I(p_2^1(x)) \in D_1$).
  \item Weak intensional equivalence "$\approx$" is defined by:\\
$\lessdot \phi \gtrdot_{\alpha}^{\beta_1}/g ~\approx~  \lessdot \psi
\gtrdot_{\alpha}^{\beta_2}/g~~~~~$
iff $~~~~~\diamondsuit  \phi[\beta_1/g(\beta_1)] \equiv \diamondsuit \psi[\beta_2/g(\beta_2)]$.\\
The symbol  $\diamondsuit = \neg \Box \neg$ is the correspondent
existential modal operator.\\
This weak equivalence is used for P2P database integration in a
number of papers
\cite{Majk03s,Majk05w,Majk06J,Majk06Om,Majk08f,Majk08rdf,Majk08in}.
\end{itemize}
Notice that we do not use the intensional equality in  our language,
thus we do not need the correspondent operator in intensional
algebra $~\A_{int}$ for the logic "necessity" modal operator
$\Box$.\\
 This
semantics is equivalent to the algebraic semantics for $\L$ in
\cite{Beal79} for the case of the conception where intensional
entities are considered to be \emph{equal} if and only if they are
\emph{necessarily equivalent}. Intensional equality is much stronger
that the standard \emph{extensional equality}  in the actual world,
just because requires the extensional equality in \emph{all}
possible worlds, in fact, if $\lessdot \phi
\gtrdot^{\beta_1}_{\alpha}/g \asymp \lessdot \psi
\gtrdot^{\beta_1}_{\alpha}/g~$ then $h(I(\lessdot A
\gtrdot^{\beta_1}_{\alpha}/g)) = h(I(\lessdot \psi
\gtrdot^{\beta_2}_{\alpha}/g))~$ for all extensionalization
functions $h \in \E$ (that is possible worlds $is^{-1}(h) \in
\widetilde{\W}$).\\
It is easy to verify that the intensional equality means that in
every possible world $w \in \widetilde{\W}$ the intensional entities
$u_1$ and $u_2$ have the same extensions.\\
Let the logic modal formula $\Box \phi[\beta_1/ g(\beta_1)]$, where
the assignment $g$ is applied only to free variables in $\beta_1$ of
a formula $\phi$ not in the list of variables in $\alpha =
(x_1,...,x_n)$, $n \geq 1$, represents a n-ary intensional concept
such that $I(\square \phi[\beta_1/ g(\beta_1)]) \in D_n$ and
$I(\phi[\beta_1/ g(\beta_1)]) = I(\lessdot \phi
\gtrdot^{\beta_1}_{\alpha}/g) \in D_n$. Then the extension of this
n-ary concept is equal to (here the mapping $necess:D_i\rightarrow
D_i$ for each $i \geq 0$
is a new operation of the intensional algebra $~\A_{int}$ in Definition \ref{def:int-algebra}):\\
$ h(I(\Box \phi[\beta_1/ g(\beta_1)]) = h(necess(I(\phi[\beta_1/ g(\beta_1)]))) =\\
 = \{(g'(x_1),...,g'(x_n))~|~\M
\models_{w,g'} \square \phi[\beta_1/ g(\beta_1)]~$ and $g' \in \D^{\V}\}$\\
$ = \{ (g'(x_1),...,g'(x_n))~|~g' \in \D^{\V}$ and $\forall w_1
((w,w_1) \in \R$ implies $\M
\models_{w_1,g'}  \phi[\beta_1/ g(\beta_1)])~\}$\\
$= \bigcap_{h_1 \in ~\E} h_1(I(\phi[\beta_1/ g(\beta_1)]))$.\\
While,\\
$ h(I(\diamondsuit \phi[\beta_1/ g(\beta_1)]) =  h(I(\neg \Box \neg \phi[\beta_1/ g(\beta_1)]) \\
= h(neg(necess(I(\neg \phi[\beta_1/ g(\beta_1)])))) \\
= \D^n \backslash h(necess(I(\neg \phi[\beta_1/ g(\beta_1)]))) \\
= \D^n \backslash (\bigcap_{h_1 \in ~\E} h_1(I(\neg \phi[\beta_1/ g(\beta_1)]))) \\
= \D^n \backslash (\bigcap_{h_1 \in ~\E} h_1(neg(I(\phi[\beta_1/ g(\beta_1)])))) \\
= \D^n \backslash (\bigcap_{h_1 \in ~\E} \D^n \backslash
h_1(I(\phi[\beta_1/ g(\beta_1)])))\\
= \bigcup_{h_1 \in ~\E} h_1(I(\phi[\beta_1/ g(\beta_1)]))$.\\
 Consequently, the concepts $\Box \phi[\beta_1/ g(\beta_1)]$ and $\diamondsuit \phi[\beta_1/ g(\beta_1)]$
are the \emph{built-in} (or rigid) concept as well, whose extensions
does not depend on possible worlds.\\
Thus, two concepts are intensionally \emph{equal}, that is,
$\lessdot \phi \gtrdot_{\alpha}^{\beta_1}/g ~\asymp~  \lessdot \psi
\gtrdot_{\alpha}^{\beta_2}/g$, iff $~h(I(\phi[\beta_1/g(\beta_1)]))
= h(I(\psi[\beta_2/g(\beta_2)]))$ for every $h$. \\Moreover, two
concepts are \emph{weakly }equivalent, that is, $\lessdot \phi
\gtrdot_{\alpha}^{\beta_1}/g ~\approx~  \lessdot \psi
\gtrdot_{\alpha}^{\beta_2}/g$, iff $~~h(I(\diamondsuit
\phi[\beta_1/g(\beta_1)])) = h(I(\diamondsuit
\psi[\beta_2/g(\beta_2)]))$.
%
\section{Application to the intensional FOL without abstraction
operator}
In the case for the intensional FOL defined in Def.
\ref{def:bealer}, without the Bealer's intensional abstraction
operator $\lessdot \gtrdot$, we obtain the syntax of the standard
FOL but with intensional semantics as presented in
\cite{Majk09FOL}.\\
Such a FOL has a well known Tarski's interpretation, defined as
follows:
\begin{itemize}
  \item  An interpretation (Tarski) $I_T$ consists in a non empty
domain
   $\D$ and a mapping that assigns to any predicate letter $p_i^k \in
   P$ a relation $R = I_T(p_i^k) \subseteq \D^k$, to any functional letter $f_i^k \in
   F$ a function $I_T(f_i^k): \D^k \rightarrow \D$, or, equivalently, its graph relation $R = I_T(f_i^k)\subseteq \D^{k+1}$ where the $k+1$-th column is
   the resulting function's value, and to each
   individual constant $c \in F$ one given element $I_T(c) \in
   \D$.\\
   Consequently, from the intensional point of view, an
   interpretation of Tarski is a possible world in the Montague's
   intensional semantics, that is $w = I_T \in \W$. The corespondent extensionalization function if $h = is(w) = is(I_T)$.\\
We define  the satisfaction for the logic formulae in $\L$ and a
given assignment $g:\V \rightarrow \D$ inductively, as follows:\\
 If a formula $\phi$ is an atomic formula $p_i^k(t_1,...,t_k)$,
then this assignment $g$ satisfies $\phi$ iff
$(g^*(t_1),...,g^*(t_k)) \in I_T(p_i^k)$;
 $~g$ satisfies $\neg \phi~$ iff it does not satisfy $\phi$;
 $~g$ satisfies $\phi \wedge \psi~$ iff $g$ satisfies $\phi$ and
$g$ satisfies $\psi$;  $~g$ satisfies $(\exists x_i)\phi~$ iff
exists an assignment $g' \in \D^{\V}$ that may differ from $g$ only
for the variable $x_i \in
\V$, and $g'$ satisfies $\phi$.\\
A formula $\phi$ is \verb"true" for a given interpretation $I_T~$
iff $~\phi$ is satisfied by every assignment $g \in \D^{\V}$. A
formula $\phi$ is \verb"valid" (i.e., tautology) iff $~\phi$ is true
for every Tarksi's interpretation $I_T \in \mathfrak{I}_T$.  An
interpretation $I_T$ is a \verb"model" of a set of formulae
$\Gamma~$ iff every formula $\phi \in \Gamma$ is true in this
interpretation. We denote by FOL$(\Gamma)$ the FOL with a set of
assumptions $\Gamma$, and by $\mathfrak{I}_T(\Gamma)$ the subset of
Tarski's interpretations that are models of $\Gamma$, with
$\mathfrak{I}_T(\emptyset) = \mathfrak{I}_T$. A formula $\phi$ is
said to be a \emph{logical consequence} of $\Gamma$, denoted by
$\Gamma \Vdash \phi$, iff $\phi$ is true in all interpretations in
$\mathfrak{I}_T(\Gamma)$. Thus, $ ~\Vdash \phi$ iff $\phi$ is a tautology.\\
 The basic set of axioms of the FOL are that of the propositional logic with two
 additional axioms: (A1) $(\forall x)(\phi \Rightarrow \psi)
 \Rightarrow (\phi \Rightarrow (\forall x)\psi)$, ($x$ does not
 occur in $\phi$ and it is not bound in $\psi$), and (A2) $(\forall
 x)\phi \Rightarrow \phi[x/t]$, (neither $x$ nor any variable in $t$ occurs bound in $\phi$).
For the FOL with identity, we need the \emph{proper} axiom (A3) $x_1
\doteq x_2 \Rightarrow (x_1 \doteq x_3 \Rightarrow x_2 \doteq
x_3)$.\\
The
 inference rules are Modus Ponens and generalization (G) "if $\phi$ is a
 theorem
 and $x$ is not bound in $\phi$, then $(\forall x)\phi$ is a
 theorem".
\end{itemize}
 The standard FOL is considered as an extensional logic
because two open-sentences with the same tuple of variables
$\phi(x_1,...,x_m)$ and $\psi(x_1,...,x_m)$ are \verb"equal" iff
they have the \emph{same extension} in a given interpretation $I_T$,
that is iff $I_T^*(\phi(x_1,...,x_m)) = I_T^*(\psi(x_1,...,x_m))$,
where $I_T^*$ is the unique extension of $I_T$ to all formulae, as
follows:\\
1. For a (closed) sentence $\phi/g$ we have that $I_T^*(\phi/g) = t$
 iff $g$ satisfies $\phi$, as recursively defined above.\\
2. For an open-sentence $\phi$ with the tuple of free variables
$(x_1,...,x_m)$ we have that $I_T^*(\phi(x_1,...,x_m)) =_{def}
\{(g(x_1),...,g(x_m))~|~ g \in \D^{\V}$ and $I_T^*(\phi/g) = t
\}$.\\
 It is easy to verify that for a formula $\phi$ with the tuple of free variables $(x_1,...,x_m)$,\\ $~I_T^*(\phi(x_1,...,x_m)/g) = t~~$ iff
 $~~(g(x_1),...,g(x_m)) \in I_T^*(\phi(x_1,...,x_m))$.\\
 This extensional \emph{equality} of virtual predicates can be generalized to the extensional \emph{equivalence}
 when both predicates $\phi, \psi$ has the same set of free variables but their
 ordering in the \emph{tuples} of free variables are not identical:
 such two virtual predicates are equivalent if the extension of the
 first is equal to the proper permutation of columns of the extension of the
 second virtual predicate. It is easy to verify that such an
 extensional equivalence corresponds to the logical equivalence denoted by $\phi \equiv \psi$.\\
 This extensional equivalence between two
relations $R_1, R_2 \in \mathfrak{R}$ with the same arity will be
denoted by $R_1 \cong R_2$, while the extensional identity will be
denoted in the standard way by $R_1 = R_2$.\\
Let $\A_{FOL} = (\L, \doteq, \top, \wedge, \neg, \exists)$ be a free
syntax algebra for "First-order logic with identity $\doteq$", with
the set $\L$ of first-order logic formulae,  with $\top$ denoting
the tautology formula (the contradiction formula is denoted by $
\neg \top$), with the set of variables in $\V$ and the domain of
values in $\D$ . It is well known that we are able to make the
extensional algebraization of the FOL by using the \emph{cylindric}
algebras \cite{HeMT71} that are the extension of Boolean algebras
with a set of binary operators for the FOL identity relations and a
set of unary algebraic operators ("projections") for each case of
FOL quantification $(\exists x)$. In what follows we will make an
analog extensional algebraization over $\mathfrak{R}$ but by
interpretation of the logic conjunction $\wedge$ by a set of
\emph{natural join} operators over relations introduced by Codd's
relational algebra \cite{Codd70,Piro82} as a kind of a predicate
calculus whose interpretations are tied to the database.
\begin{coro} \label{coro:intensemant} \textsc{Extensional FOL
 semantics:}\\
Let us define the extensional relational algebra  for the FOL by,\\
$\A_{\mathfrak{R}} = (\mathfrak{R}, R_=, \{<>\}, \{\bowtie_{S}\}_{ S
\in \P(\mathbb{N}^2)}, \sim, \{\pi_{-n}\}_{n \in \mathbb{N}})$,
\\where $ \{<>\} \in \mathfrak{R}$ is the algebraic value
correspondent to the logic truth, and $R_=$ is the binary relation
for extensionally equal elements.
We will use '$=$' for the extensional identity for relations in $\mathfrak{R}$.\\
Then, for any Tarski's interpretation $I_T$ its unique extension to
all formulae $I_T^*:\L \rightarrow \mathfrak{R}$ is also the
homomorphism $I_T^*:\A_{FOL} \rightarrow \A_{\mathfrak{R}}$ from the
free syntax FOL algebra into this extensional relational algebra.
\end{coro}
\textbf{Proof:} In \cite{Majk09FOL}\\$\square$\\
 Consequently, we obtain the following Intensional/extensional FOL semantics
 \cite{Majk09FOL}:\\
For any Tarski's interpretation $I_T$ of the FOL, the following
 diagram of homomorphisms commutes,
\begin{diagram}
    &    & \A_{int}~ (concepts/meaning) & &\\
 & \ruTo^{I ~(intensional~interpret.)} & \frac{Frege/Russell}{semantics}  &\rdTo^{h ~(extensionalization)} &\\
 \A_{FOL}~(syntax)  &    & \rTo_{I_T^*~(Tarski's ~interpretation)} && \A_{\mathfrak{R}} ~(denotation)   \\
\end{diagram}
where $h = is(w)$ where $w = I_T \in \W$ is the explicit possible
world (extensional Tarski's interpretation).\\
This homomorphic diagram formally express the fusion of Frege's and
Russell's semantics \cite{Freg92,Russe05,WhRus10} of meaning and
denotation of the FOL language, and renders mathematically correct
the definition of what we call an "intuitive notion of
intensionality", in terms of which a language is intensional if
denotation is distinguished from sense: that is, if both a
denotation and sense is ascribed to its expressions. This notion is
simply adopted from Frege's contribution (without its infinite
sense-hierarchy, avoided by Russell's approach where there is only
one meaning relation, one fundamental relation between words and
things, here represented by one fixed intensional interpretation
$I$), where the sense contains mode of presentation (here described
algebraically as an algebra of concepts (intensions) $\A_{int}$, and
where sense determines denotation for any given extensionalization
function $h$ (correspondent to a given Traski's interpretaion
$I_T$). More about the relationships between Frege's and Russell's
theories of meaning may be found in the Chapter 7,
"Extensionality and Meaning", in \cite{Beal82}.\\
As noted by Gottlob Frege and Rudolf Carnap (he uses terms
Intension/extension in the place of Frege's terms sense/denotation
\cite{Carn47}), the two logic formulae with the same denotation
(i.e., the same extension for a given Tarski's interpretation $I_T$)
need not have the same sense (intension), thus such co-denotational
expressions are not
\emph{substitutable} in general.\\
In fact there is exactly \emph{one} sense (meaning) of a given logic
formula in $\L$, defined by the uniquely fixed intensional
interpretation $I$, and \emph{a set} of possible denotations
(extensions) each determined by a given Tarski's interpretation of
the FOL as follows from Definition \ref{def:intensemant},
\begin{center}
$~~~ \L ~\longrightarrow_I~ \D ~\Longrightarrow_{h = is(I_T) \& I_T
\in ~\W = \mathfrak{I}_T(\Gamma)}~ \mathfrak{R}$.
\end{center}
Often 'intension' has been used exclusively in connection with
possible worlds semantics, however, here we use (as many others; as
Bealer for example) 'intension' in a more wide sense, that is as an
\emph{algebraic expression} in the intensional algebra of meanings
(concepts) $\A_{int}$ which represents the structural composition of
more complex concepts (meanings) from the given set of atomic
meanings. Consequently, not only the denotation (extension) is
compositional,
but also the meaning (intension) is compositional.\\

%
\section{Conclusion}
Semantics is the theory concerning the fundamental relations between
words and things. In Tarskian semantics of the FOL one defines what
it takes for a sentence in a language to be true relative to a
model. This puts one in a position to define what it takes for a
sentence in a language to be valid. Tarskian semantics often proves
quite useful in logic. Despite this, Tarskian semantics neglects
meaning, as if truth in language were autonomous. Because of that
the Tarskian theory of truth becomes inessential to the semantics
for more expressive logics, or more 'natural' languages.\\
Both, Montague's and Bealer's approaches were useful for this
investigation of the intensional FOL with intensional abstraction
operator, but the first is not adequate and explains why we adopted
two-step intensional semantics (intensional interpretation
with the set of extensionalization functions). \\
At the end of this work we defined an extensional algebra for the
FOL (different from standard cylindric algebras), and the
commutative homomorphic diagram that express the generalization of
the Tarskian theory of truth for the FOL into the Frege/Russell's
theory of meaning.


\bibliographystyle{IEEEbib}
\bibliography{medium-string,krdb,mydb}



%
\end{document}